\documentclass[english]{article}

\usepackage{epsfig}

\begin{document}

\vspace{4.8cm}

\begin{center}
{\Large {\bf Particle production in high energy collisions}} 
\vspace{0.8cm}

\normalsize{G. Giacomelli \\ 
{\small
Physics Department of the University of Bologna and INFN Sezione 
di Bologna, I-40127 Bologna, Italy} \\
Email : giacomelli@bo.infn.it}

%\vspace{1cm}

%{\large \bf In honour of Dumitru B. Ion}

\end{center}

\vspace{0.5cm}

{\bf Abstract.} {\normalsize 
A short historical review is made of charged particle production at high energy proton synchrotrons and at $pp$ and $\overline{p}p$ colliders. The review concerns mainly low $p_{t}$ processes, including diffraction processes, and fragmentation of nuclei in nucleus-nucleus collisions. A short recollection is made of the first studies of high $p_{t}$ processes. Conclusions and perspectives follow.}

\section{Introduction}

At the end of the $20^{th}$ century several high-quality secondary beams of charged particles were produced at proton 
synchrotrons of increasing energy: at the PS and SPS at CERN, the AGS at BNL, the 70 GeV proton synchrotron at IHEP Serpukhov, and at the separated function proton synchrotron at Fermilab. The secondary charged hadron beams contained the six stable or quasi stable charged particles $\pi^{\pm}$, $K^{\pm}$, $p$, $\bar{p}$, and very small backgrounds. After several experiments on hadron production \cite{baker, bez, bertin}, many experiments on total hadron-hadron ($hh$) cross sections \cite{Giac, cool, allaby, carroll2} and hadron-nucleus absorption cross sections \cite{binon} were performed, using the transmission method in ``good geometry". 
    In some high-intensity beams the production of the charged hadrons was studied and is being studied in more detail \cite{baldo}, and several particle searches were made, which also lead to the study of $\bar{d}$, $\bar{t}$, and $\overline{He^3}$ production \cite{bez}.
 
    In order to reach higher energies, hadron-hadron colliders were build. The first was the ISR $pp$ and $\bar{p}p$ collider at CERN, which allowed to reach a maximum c.m. energy of 63 GeV; then the $S\bar{p}pS$ collider at CERN allowed $\bar{p}p$ collisions up to 600 and 900 GeV c.m. energies;  the Fermilab Tevatron $\bar{p}p$ collider allowed c.m. energies up to 1.8 and 2.0 TeV. Finally the RHIC collider at BNL allows $pp$ and gold-gold nuclear collisions at c.m. energies around 200 GeV/nucleon. 
    
    Some of the experiments performed at the colliders were relatively simple  dedicated experiments, like those for total $hh$ cross sections and elastic scattering measurements \cite{amaldi, bozzo, amos, avila}, and single arm spectrometers for the study of low $p_{t}$ inelastic collisions \cite{antinucci}. But then followed elaborated general purpose detectors \cite{brea, gg3}: the new ones all had a central detector, an electromagnetic calorimeter, a hadron calorimeter and a muon detector, and the experiments had a very large number of electronic channels and needed hundreds of physicists and engineers \cite{abe}, and the sociology of the experiments changed considerably \cite{ggi}. Experiments at the RHIC collider use a variety of detectors, most of which are refined and complex detectors with many electronic channels \cite{bnl}.
 
\begin{figure}[!t]
\begin{center}
\vspace{-1cm}
{\centering\resizebox*{!}{4.6cm}{\includegraphics{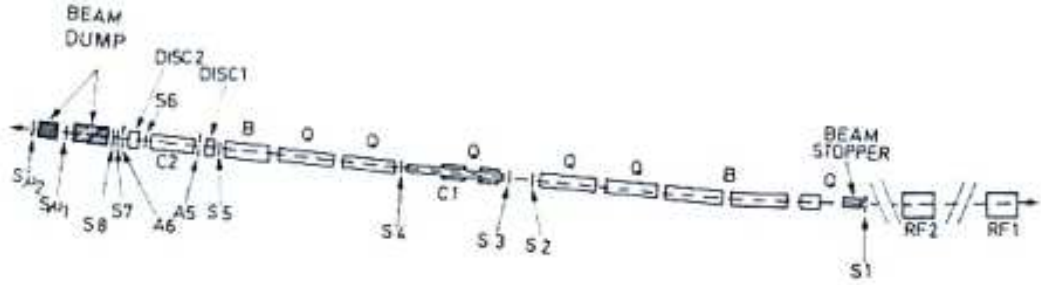}}}
\vspace{-0.6cm}
\begin{quote} 
\caption{\small Layout of the $S_1$, RF separated beam at the CERN SPS. $S_1$-$S_3$, $A_5$, $A_6$, $S_{\mu1}$, $S_{\mu2}$ were scintillation counters. D's were Cherenkov gas differential counters of the DISC type, C's were threshold gas Cherenkov counters; A and B were quadrupoles and bending magnets.}
\label {fig:aaa}
\vspace{-0.5cm}
\end{quote}
\end{center}
\end{figure}
    
   The Large Hadron Collider (LHC) at CERN will allow $pp$ collisions up to c.m. energies of 14 TeV and lead-lead collisions up to c.m. energies of $\sqrt{s_{NN}}$=5.5 TeV/nucleon. Two general purpose detectors (ATLAS and CMS) are designed to investigate the largest range of physics, while one experiment (ALICE), focuses on the search for the Quark-Gluon Plasma (QGP) \cite{link}. These experiments use large, elaborate detectors with an incredibly large number of electronic channels and new sophisticated and complex computers; each experiment was constructed and will be operated by thousands of physicists and engineers coming from Universities and laboratories from all over the world. Other specialized detectors are designed for b-physics (LHCb), total cross section (TOTEM) and forward physics (LHCf) \cite{link}. LHC will open up a completely new and unexplored energy region.
   
   Higher energies were and still are obtained with cosmic rays \cite{yodh}; and also cosmic ray experiments are becoming very large, like the Auger experiment \cite{auger}.
   
   In this paper we shall mainly concentrate on particle production at high energies and on some features of low $p_{t}$ inelastic collisions. 

\section{Measurements of particle production at accelerators}

Secondary charged beams at proton synchrotrons were momentum selected by a number of magnetic dipoles, focused by several quadrupoles and defined by collimators. Wanted beam particles were further defined by scintillation counters and identified by precise differential Cherenkov counters. Sufficient $\pi^{+}$, $K^{+}$ separation was easy to achieve at low energies, and required very selective Cherenkov counters with corrected optics at the highest energies. Contaminations of unwanted particles were kept  at a very low level ($<$$0.1\%$); electromagnetic radiations were measured with a lead glass Cherenkov counter ($e^{+}$, $e^{-}$, $\gamma$); muons were identified by their ability to pass through several meters of steel.

   A special high intensity beam made at the CERN SPS is schematically shown  in Fig. 1 and illustrated in section 4. Normally used beams were similar to that shown in Fig. \ref{fig:aaa}, but were simpler, of lower intensity and had no RF separators. 
   
\begin{figure}[!t]
 \centering
 \vspace{-1cm}
 {\centering\resizebox*{!}{7cm}{\includegraphics{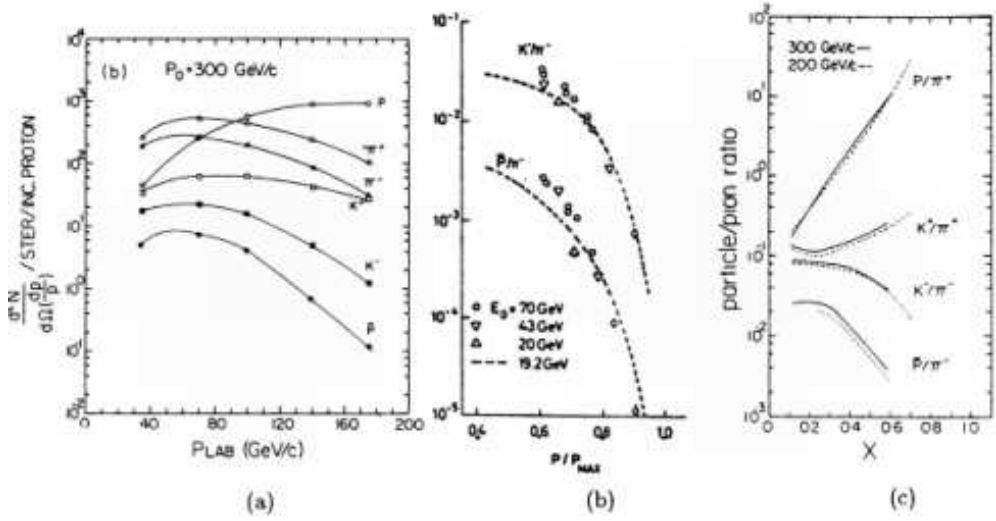}}\par}
\begin{quote}
 \caption{\small (a) Production cross sections of the six long-lived charged hadrons plotted vs lab momentum (incident proton beam $p_{lab}$=300 GeV/c); (b) (c )particle ratios vs $p/p_{max}$ at $p_{lab}$=70 GeV at IHEP-Serpukhov. } 
\vspace{-1cm}
\label{fig:long}
\end{quote}
 \end{figure}

   Fig. \ref{fig:long}a shows the production cross sections of the six charged long lived hadrons plotted versus beam momentum at Fermilab for an incident proton beam of 300 GeV/c. Fig. \ref{fig:long}b shows particle ratios vs $p/p_{max}$ at the 70 GeV IHEP proton synchrotron. It is interesting to note that the ratios are essentially independent of beam energies. Fig. \ref{fig:long}c shows particle ratios, $p/\pi^{+}$, $K^{+}/\pi^{+}$, $K^{-}/\pi^{-}$, $\overline{K}/\pi^{-}$, measured at Fermilab.     
   
   Recently, the main motivation for new measurements of pion yields is neutrino physics, in particular for a quantitative design of a proton driver for a future neutrino factory, for a substantial improvement in the calculation of the atmospheric neutrino fluxes and the measurement of particle yields for accelerator neutrino measurements \cite{baldo}. The HARP experiment at CERN obtained new extensive data, as shown in Fig. \ref{fig:fig1010}.      
   
\section{ Inelastic low {\boldmath $p_T$} processes}

 \begin{figure}[!b]
\begin{center}
{\centering\resizebox*{!}{7cm}{\includegraphics{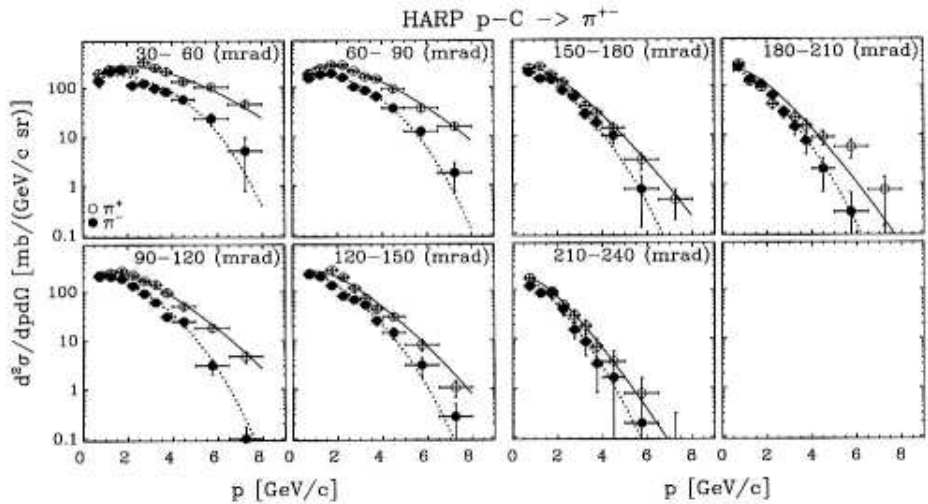}}}     
\begin{quote} 
\caption{\small Measurements of the double differential production cross section of positive (open circles) and negative (full circles) pions from 12 GeV/c protons on carbon as a function of pion momentum $p$ in the lab. frame \cite{baldo}. }
\label {fig:fig1010}
\end{quote}
\end{center}
 \end{figure}

\begin{figure}[!t]
\begin{center}
\vspace{-1cm}
{\centering\resizebox*{!}{4.4 cm}{\includegraphics{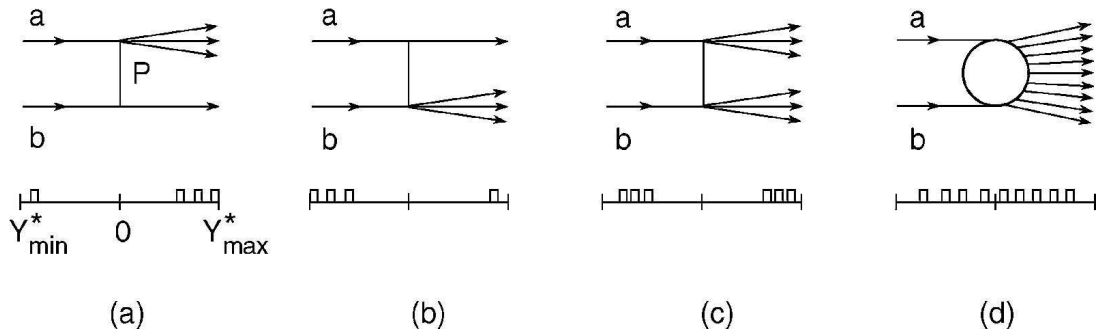}}}     
\begin{quote} 
\caption{\small Pictorial description of inelastic, small $p_{t}$ processes with characteristic rapidity distributions.
P indicates Pomeron exchange. (a) Fragmentation of the beam particle a; (b) fragmentation
of the target particle b; (c) double fragmentation of a and b; (d) inelastic non diffractive collisions.}
\label {fig:fig2}
\end{quote}
\vspace{-1cm}
\end{center}
 \end{figure}

Fig. \ref{fig:fig2} gives a pictorial representation of the dominant inelastic processes in $hh$ collisions at low $p_{t}$: single and double diffraction dissociation and inelastic processes, which concentrate in the forward direction; this  is also the important direction for cosmic ray data. The total $\bar{p}p$ cross section may be written as

\begin{equation}
\sigma_{tot}= \sigma_{el} + \sigma_{inel} = \sigma_{el} + \sigma_{sd} + \bar{\sigma}_{sd} + \sigma_{dd} + \sigma_{nd}                                                                                                                                            
\end{equation}\\
where $\sigma_{el}$ is the elastic cross section, $\sigma_{sd}$ is the single diffractive cross section for the incoming proton, $\bar{\sigma}_{sd}$ is the single diffractive cross section for the incoming antiproton ($\bar{\sigma}_{sd}$ $\simeq$ $\sigma_{sd}$), $\sigma_{dd}$ is the double diffractive cross section and $\sigma_{nd}$ is the non diffractive part of the inelastic cross section. Most of the non diffractive cross section concerns hadrons emitted with low transverse momenta ($low$ $p_{t}$ $physics$) with properties which change slowly with c.m. energy ($\it{ln}$${s}$ $physics$) 
\cite{gg3}.

   The diffractive cross section is difficult to measure and it is not too well known experimentally, as indicated in Fig. \ref{fig:fig1010}.    
                                                                                                                                                                                                                                                                                                                                                                                                                    

   A small part of the non diffractive cross section is due to central collisions  between the two colliding particles  and gives rise to high $p_{t}$ jets of particles emitted at large angles ($large$ $p_{t}$ $physics$); this contribution increases with increasing energies and eventually becomes dominant.
   
    Charged multiplicities were obtained by interpolations of measurements with single arm spectrometers \cite{antinucci} and by more direct measurements with large acceptance ($\sim$4$\pi$) detectors  \cite{brea}.
   
   The average number of charged hadrons, $\langle n_{ch} \rangle$, produced 
in high energy collisions increases with increasing c.m. energy $\sqrt{s}$, as shown in Fig. \ref{fig:fig3}a for the charged multiplicities in $\bar{p}p$ and $pp$ collisions \cite{antinucci, brea}. The data may be fitted to a power law dependence in $\ln{s}$ of the type:
  
  \begin{figure}[!b]
\centering
{\centering\resizebox*{!}{6cm}{\includegraphics{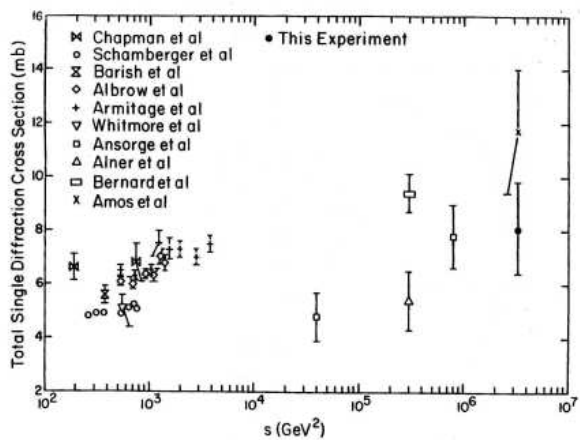}}}
\begin{quote} 
\caption{\small The total single diffraction cross section measured by 
different experiments \cite{amos}.}
\label {fig:fig92}
\end{quote}
\end{figure}

\begin{figure}[!t]
\begin{center}
\vspace{-1cm}
{\centering\resizebox*{!}{7.6 cm}{\includegraphics{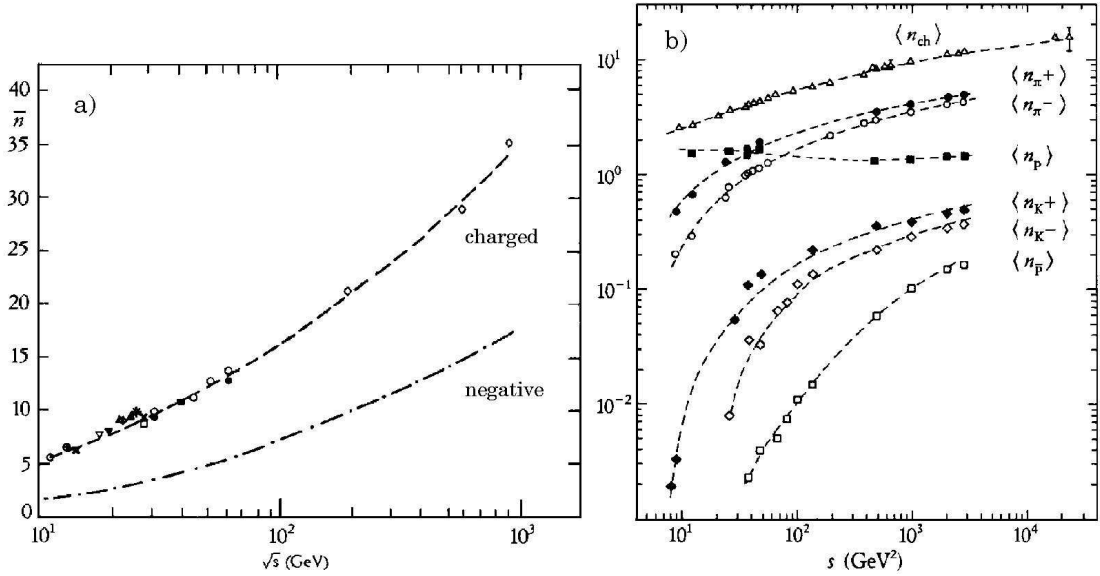}}}
\begin{quote} 
\caption{\small (a) Average charged multiplicities $\bar{n}$ for $\bar{p}p$ and $pp$ collisions $vs$. $E_{cm}$=$\sqrt{s}$. (b) Average
number of $\pi^{\pm}$, $K^{\pm}$, $p$ and $\bar{p}$ produced in $pp$ collisions at the CERN ISR for c. m. energies up to 63 GeV.}
\label {fig:fig3}
\vspace{-1cm}
\end{quote}
\end{center}
 \end{figure}

\begin{figure}[!b]
\begin{center}
\vspace{-0.4cm}
{\centering\resizebox*{!}{6 cm}{\includegraphics{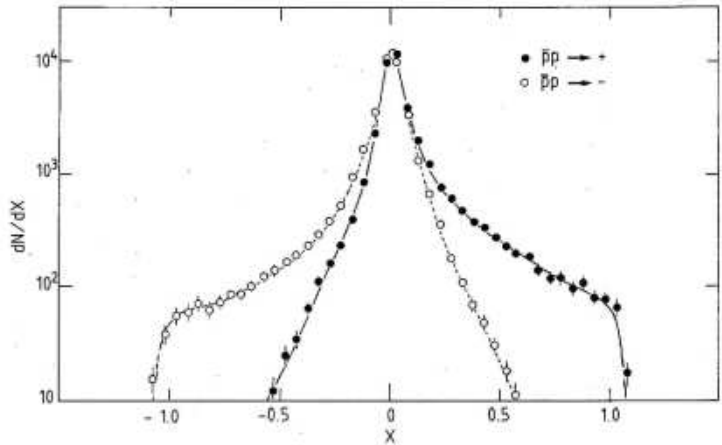}}}
\begin{quote} 
\caption{\small $x$-distributions of positive and negative particles produced 
in $\bar{p}p$ interactions at $\sqrt{s}=$53 GeV \cite{brea}. Positive values of $x$ correspond to the direction of the incoming protons. The slight differences between leading particles for $x$$>$$0$ and $x$$<$$0$ are due to the acceptance of the apparatus. The lines are only meant to guide the eye.}
\label {fig:fig4}
\vspace{-0.5cm}
\end{quote}
\end{center}
 \end{figure}

\begin{equation}
\langle n \rangle = A + B \ln s + C \ln^{2} s \simeq 3.6 - 0.45 \ln s + 0.2 \ln^{2} s
\end{equation}\\
In $\bar{p}p$ collision at $\sqrt{s}$=1.8 TeV are produced on average about 40 charged particles and 20 neutral particles in each collision. Fig. \ref{fig:fig3}b shows that most of the produced particles are pions, followed by kaons; it is also visible the so called $leading$ $effect$, which is connected with the emission of relatively many high energy protons in the incoming proton direction (antiprotons in the incoming direction of antiprotons), see also Fig. \ref{fig:fig4} \cite{gg3, abe}. The computations of charged multiplicities at higher  energies are based on Monte Carlo methods, which have considerable uncertainties \cite{mor}. Also the measurements are not easy. It is instead easier to measure and calculate the multiplicities from quark and gluon jets at large $p_{t}$ \cite{daga}. It is interesting to recall that gluon jets yield larger charged multiplicities than quark jets.  

   At the LHC, for $pp$ collisions at $\sqrt{s}\sim$14 TeV, one expects the production of 70-90 charged particles per collision \cite{mor}.
   
     The average $p_{t}$ of the produced particles increases slowly with $\sqrt{s}$: it was $\sim$0.36 GeV/c in 20$<$$\sqrt{s}$$<$100 GeV and it increased to $\sim$0.46 GeV/c at $\sqrt{s}$=1.8 TeV. The simplest interpretation of these features is in terms of thermodynamic models \cite{gg3}.

   It may be stressed that some of the main qualitative features of particle production at high energies and low $p_{t}$ are easily observed in the Feynman $x$ distribution ($x$=$p_{l}/p_{l\,max}$ in the c.m. system). Fig. \ref{fig:fig4} shows the production of positive and negative particles in $\bar{p}p$ interactions at $\sqrt{s}$=53 GeV plotted versus $x$: for $x$$>$$0$ (the direction of the incoming protons) the distribution for positive particles contains two components, fragmentation products for $x$$>$$0.3$ and centrally produced particles for $x$$<$$0.3$. The distribution for negative particles on the other hand has only one component, which is typical of centrally produced particles, i.e. with a large concentration of particles at $x$$=$$0$, rapid fall off at larger $x$, and with very few particles for $|x|$$>$$0.3$ \cite{brea, gg3}. In the region for $x$$<$$0$ the negative particles are leading particles, following the charge of the incident antiproton, while positive particles are the produced ones. For $pp$ collisions the situation is symmetric around $x$$=$$0$ and the distributions show leading effects for the positive particles only.  
 
 \begin{figure*}[t]
\begin{minipage}{0.5\linewidth}
\vspace{-2.4cm}
\hspace{-1cm}
{\centering\resizebox*{!}{5cm}{\includegraphics{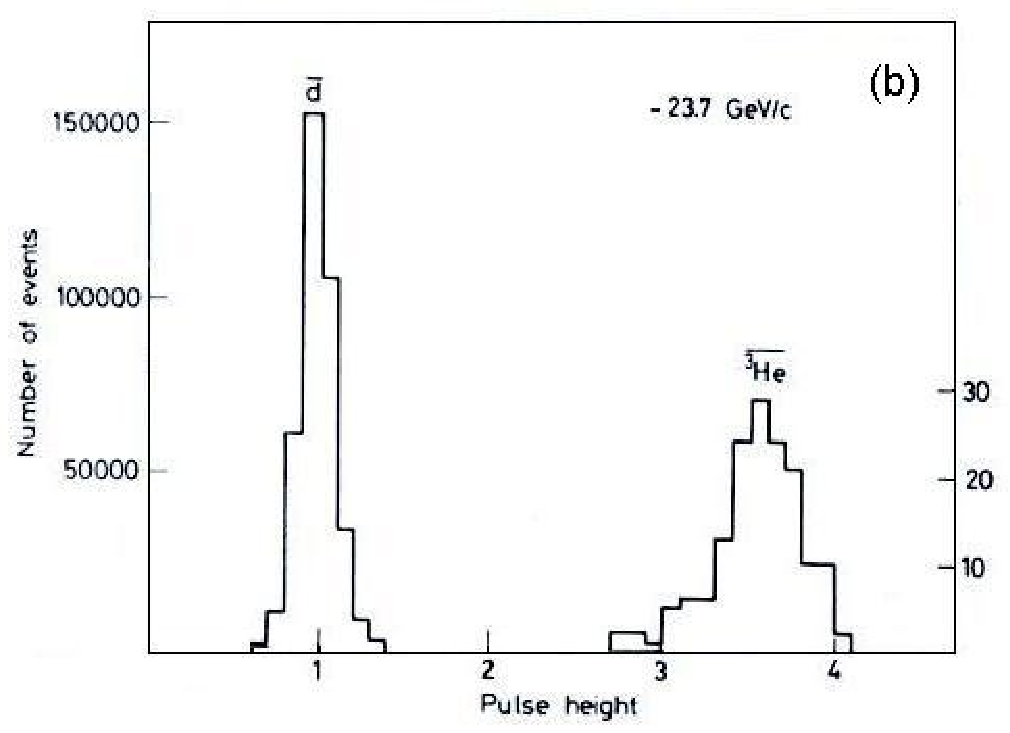}}}
\hspace{-2cm}
{\centering\resizebox*{!}{5cm}{\includegraphics{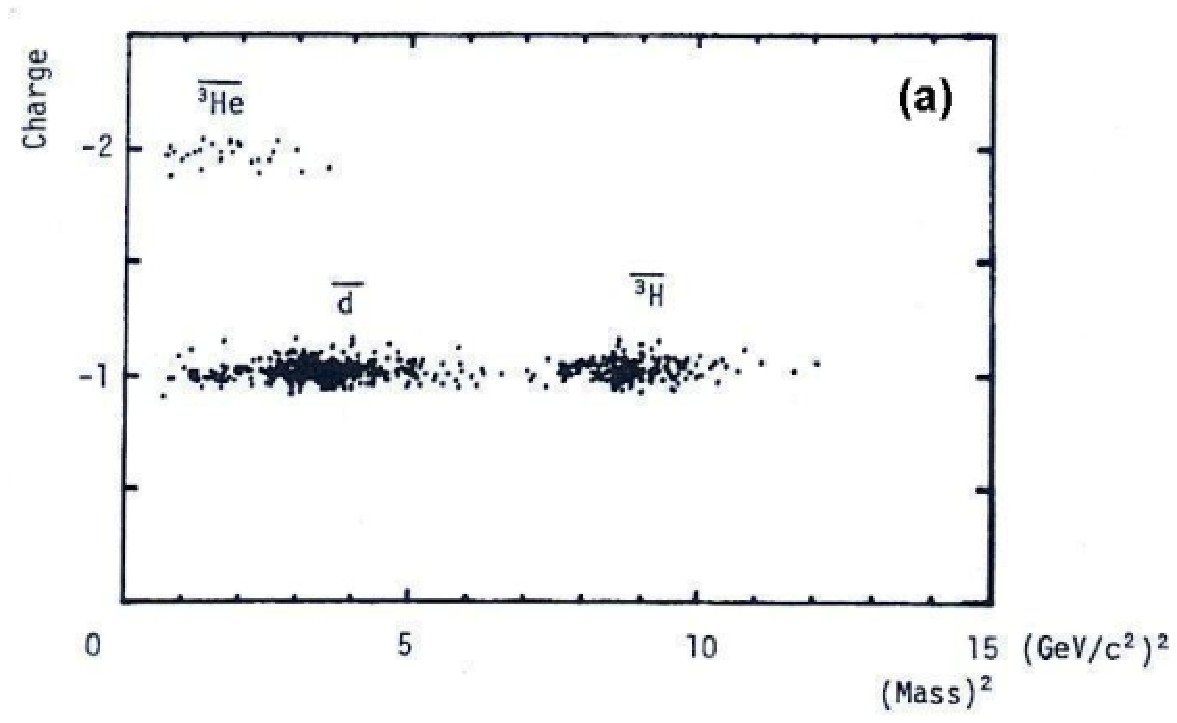}}\par}
\caption{\small (a) Scatter diagram of $charge$ versus $(mass/charge)^{2}$. $\overline{He^3}$, $\overline{t}$ and $\overline{d}$ events are clearly seen. (b) Average pulse height distribution for $\overline{d}$ and $\overline{He^3}$. }
\label{fig:zenith}
\end{minipage}\hfill
\begin{minipage}{0.45\linewidth}
{\centering\resizebox*{!}{8cm}{\includegraphics{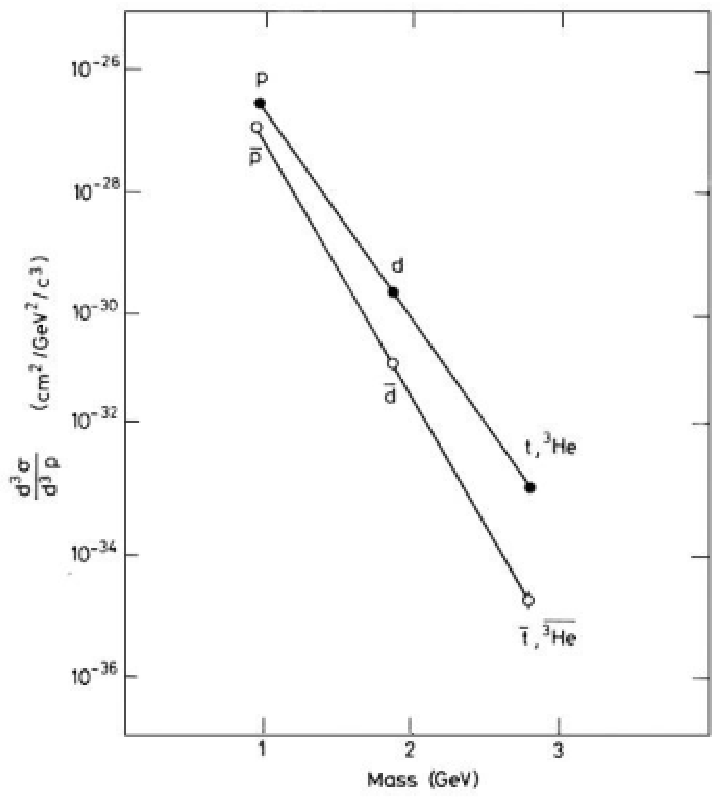}}\par}
\caption{\small Invariant cross sections for the inclusive productions of $p$, $d$, $t$, and $\overline{p}$, $\overline{d}$, $\overline{t}$, $\overline{He^3}$
 plotted versus mass. The lines are fits to an exponential form.} 
\label{fig:le}
\end{minipage}
\end{figure*}

\section{Production of antinuclei in a high intensity separated beam}  

At the CERN SPS a special high intensity radiofrequency (RF) separated beam was constructed in order to study particle production  in the very forward direction, search for new massive particles with charges larger than $|2/3|e$ and to study antinuclei (several hundred antitritons and $\overline{He^3}$ were recorded, see Fig. \ref{fig:zenith}). Fig. \ref{fig:aaa} shows the layout of the beam: the primary proton beam of $200 \div 240$ GeV/c came from the right, interacted in a beryllium or aluminum target producing a secondary beam of $6 \div 40$ GeV/c which passed through  two RF separators which removed the high velocity particles and thus allowed acceptances of more than $2 \times 10^7$ particles per burst produced at the target, while keeping the numbers of particles in the detectors at the $\sim$$10^6$ level. Particle identification was made with differential Cherenkov counters (DISC) and with time-of-flight counters, each of which had a time accuracy of 0.15 ns; particle charges were measured from the energy losses in scintillators.

The production cross sections of charged pions, kaons, protons and 
antiprotons were obtained from pressure curves with one of the DISC 
counters; in these case no use was made of the RF separators and secondary 
beams were at relatively low intensities. The search for rare particles was 
made at each secondary momentum with the RF separators on: the events not 
vetoed by a threshold Cherenkov nor by a DISC counter (set on antideuterons) 
triggered the apparatus and were recorded for further analyses. Examples of 
measured rare antinuclei are shown in Fig. \ref{fig:zenith}. The mass 
dependence of the invariant cross sections for $p$, $d$, $t$, $He^3$ and 
$\overline{p}$, $\overline{d}$, $\overline{t}$, $\overline{He^3}$ production 
is shown in Fig. \ref{fig:le}. Notice the strong decrease of
the production cross section with increasing nuclei (antinuclei) mass.
No new particles with integer or fractional 
charge were observed  and upper limits were established at the level of 
$<$$10^{-11}$ of the production of known particles, namely pions and kaons.

\section{Hadron-hadron and hadron-nuclei cross sections}

At fixed target proton synchrotron accelerators (at BNL, IHEP and Fermilab) the total cross sections of charged hadrons were measured with the transmission method in good geometry, with relative precisions smaller than 1$\%$ and systematic scale errors of 1-2$\%$ \cite{cool, allaby, carroll2}. 

   The measurements of the total cross sections at the $\bar{p}p$ and $pp$ colliders required the development of new experimental techniques: the scattering of particles was measured at very small angles, with detectors positioned  in re-entrant containers (``Roman pots") located very close to the circulating beams and far away from the interaction point. The combinations of statistical and systematic uncertainties are $\geq$10$\%$, with the exception of the CERN ISR, where the luminosity was measured accurately by the Van der Meer method of displacing vertically the beams \cite{amaldi, bozzo, amos, avila}.
   
Above the resonance region the total $hh$ cross sections decrease, reach a minimum and then increase with increasing energy    
(the $K$+$p$  total cross section was already increasing at IHEP-Serpukhov energies \cite{allaby}). The difference between the $\bar{x}p$ and $xp$ total cross sections decreases with increasing energy.

    At the highest energies there are only cosmic ray (CR) data \cite{yodh}; LHC and the new large CR experiments will improve the experimental situation at these very large energies \cite{auger}.

   As byproducts of $hh$ total cross section measurements, the absorption cross sections of charged pions, charged kaons, protons and antiprotons were measured on several target nuclei, for example Li, C, Al, Cu, Sn and Pb \cite{binon}. The energy dependence of these cross sections tends to follow that of the $hh$ cross sections, but is somewhat slower. 
   The data at each energy were fitted to the simple expression
                                                                                                                                                          
\begin{equation}
\sigma_{abs}(A) = \sigma_{0} A^{\alpha}
\end{equation} \\                                                                                                                                                           
where A is the atomic weight of the target nucleus. At very high energies the parameter $\sigma_0$ increases with energy; the parameter $\alpha$ decreases with increasing energy, with a tendency to go towards the value 2/3 for large values of $\sigma_{hp}$, as would be expected for an opaque nucleus \cite{binon, gg3}. The relations between $hh$ and h-nuclei cross sections are in most cases given in the context of the Glauber theory \cite{glauber}: inelastic scattering is treated in the shadowed single collision approximation at small momentum transfers, in the multiple collision approximation at large momentum transfers.

\begin{figure}[t!]
\begin{center}
\centering
{\centering\resizebox*{!}{5cm}{\includegraphics{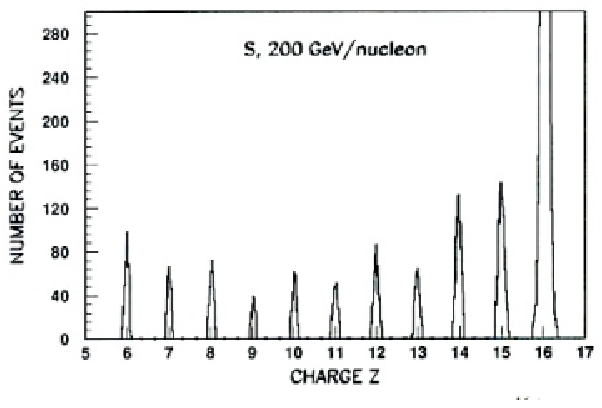}}}
\hspace{0.1cm}
{\centering\resizebox*{!}{5cm}{\includegraphics{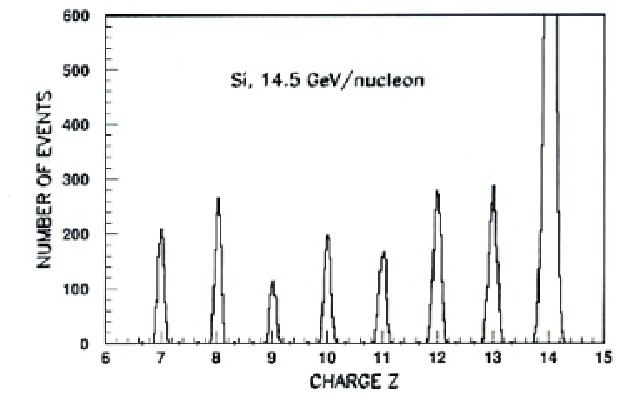}}\par}
\begin{quote}
\caption{\small Charge distribution of (a) 200 GeV/nucleon $S^{16+}$ ions and their fragments and (b) of 14.5 GeV/nucleon and their fragmentation measured with CR39 NTDs. In both cases the beam and fragments were measured in at least 10 CR39 adjacent detector sheets.} 
\label{fig:illustration2}
\vspace{-0.5cm}
\end{quote}
\end{center}
\end{figure}

\section{Fragmentation of nuclei in nucleus-nucleus collisions} 

The fragmentation cross sections of various high energy nuclei on different nuclear targets were often measured using Nuclear Track Detectors (NTDs) \cite{ccec}. These measurements are of interest for nuclear physics and also for a number of applications, like cancer therapy, evaluation of the doses received by astronauts from cosmic rays \footnote{Cosmic rays are present everywhere in space: one can say that ``space is radioactive" because of the presence of high energy
charged cosmic rays. For future long range space explorations it is important to establish the fragmentation properties of heavy ions and the radioactive doses received by astronauts.}, etc. Fig. \ref{fig:illustration2}a shows the charge distribution obtained with 200 GeV/nucleon S$^{16+}$ ions and their fragments produced in a copper target. Fig. \ref{fig:illustration2}b shows the charge distribution of fragments from Si$^{14+}$ ions of 14.5 GeV/nucleon. 
Notice in both cases the very good charge resolution for each peak obtained via the measurements of the nuclear fragments in $\sim$10 successive layers of CR39 NTDs (the resolution improves linearly  as the inverse of the square root of the number of measurements). Notice also the absence of nuclear fragments with fractional charge. One also observes the even/odd effect on the produced fragments (the height of a $Z$=even peak is higher than those of the close by peaks with $Z$=odd). 

\begin{figure}[!t]
\begin{center}
{\centering\resizebox*{!}{3.3cm}{\includegraphics{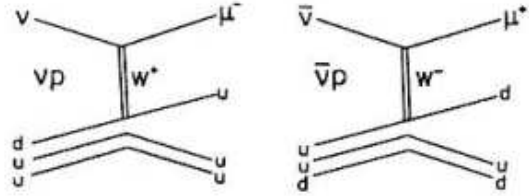}}}
\begin{quote} 
\caption{\small Dominant mechanisms for Charged Current $neutrino-proton$ 
and $antineutrino-proton$ scattering in the quark parton model.}
\label {fig:fig1111}
\vspace{-0.5cm}
\end{quote}
\end{center}
 \end{figure}

The word fragmentation is also used in other contexts, for example in 
charged current (CC) neutrino-nucleon interactions. In the quark parton model the dominant CC mechanism consists in the interaction of a $\nu_\mu$ ($\overline{\nu_\mu}$) with a $d(u)$ valence quark which is converted into a $u(d)$ quark that subsequently is ``dressed up'' by quarks and antiquarks created in the color field to form a number of particles moving in the forward direction in the c.m. system ({\it{beam fragmentation}}). In the same way the remaining diquark system ($uu$ in $\nu_p$, $ud$ in $\overline{\nu_p}$) yields backward going hadrons among which a baryon ({\it{target fragmentation}}): see Fig. \ref{fig:fig1111} \cite{Allasia}.

 \begin{figure}[!b]
\begin{center}
\hspace{-1cm}
\centering
{\centering\resizebox*{!}{5.5cm}{\includegraphics{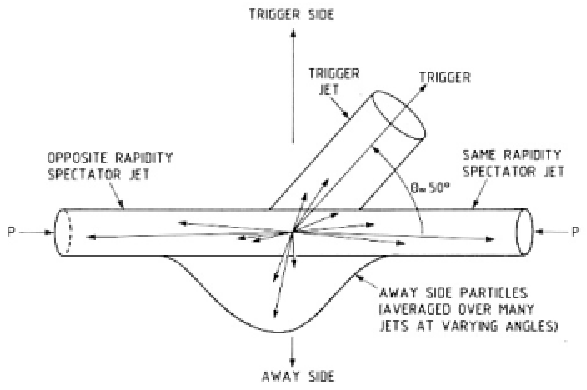}}}
\hspace{1cm}
{\centering\resizebox*{!}{5.5cm}{\includegraphics{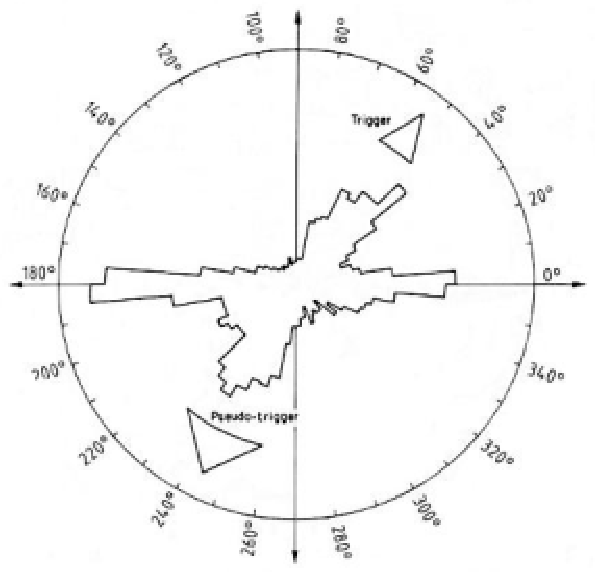}}\par}
\begin{quote}
\caption{\small (a) Schematic representation and nomenclature for hard parton-parton collisions at the SFM detector at the CERN ISR. (b) Deep inelastic event producing a charged particle with $p_{t}$$>$6 GeV/c and an additional pseudo trigger particle with $p_{t}$$>$2 GeV/c. Projection into the scattering plane of the normalized momentum flow as a function of the polar angle $\theta$ \cite{brea}. } 
\label{fig:illu}
\vspace{-0.5cm}
\end{quote}
\end{center}
\end{figure}

\section{Large {\boldmath $p_t$} events. Early analyses}

Some large $p_t$ events were first studied at the CERN ISR, where most of the effort and the cross sections measured involved low $p_{t}$ events. The Split Field Magnet (SFM) was probably the first general purpose large apparatus of high energy physics, with a complex magnetic field and a track detector covering most of the 4$\pi$ solid angle \cite{brea, gg3}. The 4 jet structure of an event was observed after analysis of the type sketched in Fig. \ref{fig:illu}. Later experiments at the proton-antiproton colliders, at the LEP positron-electron collider, at the HERA positron-proton collider, at the RHIC and soon at the LHC are mostly general purpose detectors, with a tracking device, an electromagnetic calorimeter, a hadron calorimeter and a muon detector, all with a cylindrical symmetry. In these new general purpose detectors the jet structure of the events is immediately evident, both because of the quality of these detectors and because they are operated at higher center of mass energies.  

\section{Experimentation at RHIC}

The Relativistic Heavy Ion Collider (RHIC) at the Brookhaven lab (BNL) is capable of accelerating and colliding (in 2 identical rings) gold ions up to c.m. energies of $\sqrt{s_{NN}}$$\sim$200 GeV/nucleon. The main purpose is to study the formation and the characteristics of the quark-gluon plasma, a state of matter believed to exist at sufficiently high energy densities \cite{ad}. If a RHIC collision produces a QGP, the plasma will quickly cool, expand and coalesce into hadrons; thus experimental physicists cannot observe directly the quark gluon plasma because its lifetime is too brief; they can study the hadrons that shower out of the collision. A collision that produces QGP will produce particles with different ratios than from a collision that does not produce QGP. In the QGP several jets should be often produced, but some of them are suppressed by re-interactions in the dense medium, which seems to behave more like a low viscosity liquid than a gas of free quarks and gluons \cite{ad}. 

   In addition to colliding heavy ions, RHIC is also able to collide deuteron-heavy ions and single protons ($pp$), at c.m. energies between 62 and 500 GeV/nucleon, and also polarized protons. The study of $pp$ collisions yields the high energy parameters, $\sigma_{tot}$, $\sigma_{el}$, slope of elastic scattering $B$, real to imaginary part of the forward scattering amplitude, etc. In the simplest quark parton model the proton spin is due to the spins of the three valence quarks, $u u d$. But one knows that dynamically the proton contains many gluons and sea quarks. Thus the explanation of the spin 1/2 of the proton must be more complex, and the study of high energy $pp$ collisions with polarized protons should provide further insight on the proton spin structure.

\section{ Conclusions and outlook}
A large amount of data is available from secondary beam surveys of the produced long lived hadrons, $\pi^{\pm}$, $K^{\pm}$, $p$ and $\overline{p}$. New particle production data were recently measured for the precise evaluation of the features of neutrino beams and of atmospheric neutrino fluxes \cite{baldo} \cite{bea}.

   A wealth of experimental information was obtained since the 1960s on high energy $hh$ and h-nuclei collisions starting with simple beams and simple apparatus and then with better beams at higher energies and more complex apparatus. Most of the data were interpreted in terms of phenomenological models, which were eventually codified in Monte Carlo programs of increasing complexity \cite{mor}.
 
  Also the amount of data on low $p_{t}$ inelastic processes is very large and also their interpretation was mainly performed with phenomenological models and Monte Carlo programs.
  
  Only the relatively small fraction of large $p_{t}$ events and of large $p_{t}$ jets of hadrons was interpreted in the context of perturbative QCD in terms of quarks and gluons.
 
  The advent of LHC should allow exploration in a new energy region, which will be performed by large refined and complex general purpose detectors, refined detectors for the search for the QGP and smaller refined detectors for low $p_{t}$ physics \cite{link}.
 
 \vspace{1.5cm}
{\Large \bf Acknowledgments}

I am grateful to many colleagues for cooperations and discussions. I thank Drs. Maddalena Errico, Roberto Giacomelli and Miriam Giorgini for discussions and technical support.

\bibliographystyle{plain}

\end{document}